\documentclass[11pt]{article}

\begin{document}
\large
\begin{center}
{\Large \bf JAVA-APPLET FOR RADIO SPECTRA ANALYSIS}\\[8mm]
{\bf A.S.Trushkina, A.G.Gubanov}\\[6mm]
{\it Astronomical Institute, St.-Petersburg State University,
Bibliotechnaja pl.2, St.-Petersburg, Petrodvoretz, Russia}\\[12mm]
\end{center}

\begin{center}
{\bf Abstract}
\end{center}

New services for Clusters of Galaxies Database created in the Astronomical Institute
of SPbU have been constructed. The detailed description of database and its content is available  
at \underline{http://www.astro.spbu.ru/CLUSTERS/}.

 The information contained in the database can be used for solution of the specific 
astronomical problems. One of these problems is the researches of radio emission spectra 
of clusters' objects. The researches can give help in solution of the problems of origin 
and evolution of extragalactic radio sources. Spectra of radio galaxies contain important 
information about radio power in the processes responsible for their activity.

 This report presents the software for the database that provides work with radio spectra 
of extragalactic sources. This client-application has been released as the Java-applet and 
thus provides a Web based interface that is supported on many operating systems. The 
facilities of the applet such as approximations of spectra measurement points and 
calculation of spectral parameters are illustrated in the report. Also we demonstrate 
new version of the application that realized on Java2 and gave access to additional 
functions such as printing.\\


\section{ Introduction}

The Web-server of Astronomical Institute of the St-Petersburg University had
opened the extended services on assignment of clusters data since 1995.
The description of Clusters of Galaxies Database and its content are available 
at \underline{http://www.astro.spbu.ru/CLUSTERS/}.

The information contained in the database can be used for solution of 
the specific astronomical problems. Two big projects - "Identifications of 
Radio Sources in Clusters of Galaxies" and "The Study of Radio Emission 
Spectra of Cluster Radio Sources" those carrying out in Astronomical 
Institute at present time - use the Clusters of Galaxies Database actively.
The second one  can give help in solution of the problems of origin and 
evolution of extragalactic radio sources.

This report presents the client-application for the database that provides
work with radio spectra of extragalactic sources. It has been released 
as a Java-applet and thus provides a Web based interface that is supported
on many operating systems. The facilities of the applet such as approximations
of spectra measurement points and calculation of spectral parameters are
illustrated in the report. Also we demonstrate last version of the
Java2 realized application that gives additional functions such as printing.

\section{ Clusters of Galaxies Database}

\subsection{Content of the database.}

The Clusters of Galaxies Database contains information about objects from 
ACO catalogue (\cite{ACO}).
For all 4076 rich clusters the Database includes information about the main 
optical and radio characteristics. 

There are many services on the server for the presentation of this database 
information, for example, the visualization of the cluster galaxies fields, 
the generation of requests to different astronomical data centers, 
the calculation of some parameters of the cluster objects.   

The Clusters of Galaxies Database has free access to these services and 
data by Internet. One of the given services is a Java applet for analysis
of radio spectrum data.

\subsection{ Spectra applet access.}

User obtains access to Clusters of Galaxies Database 
(\underline{http://www.astro.spbu.ru/CLUSTERS/}) by Internet-browser 
and requests necessary data by the name of the cluster.

The server processes the request and prepares data that user needs.
The client receives hypertext document containing links to the
various services and data. In particular it has the link
\underline{[radio spectra 1.2 or 1.1]} to a document with Java-applet call.
Following this link the client browser requests Java-applet from 
the server and runs it.

At first the applet loads data from the Database. On theirs receiving 
user gets capabilities to work with a graph of radio source spectrum,
to calculate its spectral characteristics and to change data
on the graph and their graphical presentation. At the beginning he can select
more comfortable sizes of user-interface widgets by menu options.

\section{Main applet functions}

Client gets the following functions to work with radio spectra.\\

{\bf The diagram functions.}\\

Spectral data are displayed as a diagram of the dependence of measurement 
flux densities with their errors on frequency of the observations. Axes of 
the diagram have logarithmic scale as usual. User can change ranges of the 
axes and put a coordinate grid. He has an essential possibility to add new 
spectral points on the graph if necessary and to delete/restore superfluous 
or non-exact measurements. In this case the approximation uses new points and 
leaves out of account the deleted points automatically. 

Also user can save diagrams as jpg-files on his computer.\\

{\bf The approximation of the spectral data.}\\

The radio spectrum measurement points are approximated by a curve for the 
calculation of spectral parameters (radio emission power, spectral index 
and others).
The curve type is defined by some general functional dependencies.
The least squared method is used for approximation of the spectral data. 
User can choose one of curve types himself. Also the applet has an automatic
option of a choice of the best approximation that has minimum deviation 
from the measurements.
The line with spectral index $\alpha = 0.8$ is used as default approximation 
for only one measurement point.
The curve values of frequency, of flux density and 
of spectral index are displayed by mouse moving along curve in the special 
{\it GraphInfo} window.\\

{\bf The calculation of the radio emission parameters.}\\

User can calculate radio luminosity for the selected radio source at
any frequency range and the spectral power of the emission at any frequency.
As is well known, the calculation of these values for extragalactic radio
sources depends on a choice of the cosmological model. So it's used the
most common formulas contained the necessary cosmological parameters, as
Hubble constant, deceleration parameter q$_{0}$, redshift and the comoving 
frames calculations. The {\it RadioEmission} window allows to user to choose 
these parameters. 
In principle the applet permits to calculate the spectral parameters for any 
distant radio source (even for radio source out of database).\\

{\bf The printing.}\\

The capability of Java2 enables to realize the printing function. The applet 
allows to user to print the graph of the radio spectrum. There is a special 
{\it PrintView} window containing the options for the preparing the graph
to print. User can choose the parameters of paper (its format,
orientation and position of graph) and change a name of graph and axes titles, 
their colors, fonts and positions.
And after diagram preparing it is possible to print 
the diagram or to save in the jpg-file.\\

{\bf The help.}\\

All functions have the detailed descriptions. The help contains a set of 
hypertext documents, concerned with each other. They are available from
the applet and from Cluster Database user interface.

\section{Applet structure}

Java as object-oriented language requires deep analysis of the application 
domain, flexible organization of the Graphical User Interface (GUI) and 
their clear interaction with each other under applet construction. During 
working and maintenance of the applet the authors design the main required 
functions and three basic versions of the radio spectral applet were realized. 
The final version of the applet uses the all past experience.

Spectra applet consists of five packages. The main package {\bf spectra} responds
in the initialization of all modules of the applet, its life cycle and stops it.
The package {\bf spectra.clusterData} contains classes to storage parameters 
of cluster and its objects and to request database information. The module 
{\bf spectra.graphics} gives a tool to the interactive work with a graph of a 
spectrum. The  {\bf spectra.emission} consists of classes used for the calculation 
the radio emission parameters. The package {\bf spectra.print} provide the functions
for the preparing and the printing a graph of a spectrum. Standard
{\bf java.awt} package is used for the building GUI of the applet.

In general there are about 50 classes in the applet. Classes of the applet 
may be used as class library to support other astronomical information
systems.\\

At present there are two versions of the applet on the server of
Astronomical Institute those are realized on JDK1.1 and Java2 
correspondingly. It was done for useful access to the 
applet because most of Internet-browsers have the built-in JDK1.1 support
still but ones require to install the plug-ins for Java2. \\

{\bf Acknowledgments.}
This work was supported by the ``Integration'' program (A0145).

\end{document}